\documentclass[fp,twocolumn]{jpsj3}
\usepackage{txfonts}
\usepackage[dvipdfmx]{graphicx}
\usepackage{amsmath,amssymb}
\usepackage{bm}
\usepackage{makeidx}
\usepackage{float}
\renewcommand{\Vec}[1]{\mbox{\boldmath$#1$}}
\title{Strongly enhanced superconductivity due to finite energy spin fluctuations induced by an incipient band : a FLEX study on the bilayer Hubbard model with vertical and diagonal interlayer hoppings}

\author{Karin Matsumoto\cite{KM}, Daisuke Ogura\cite{DO}, and Kazuhiko Kuroki}
\inst{Department of Physics, Osaka University, 1-1 Machikaneyama, 
Toyonaka, Osaka 560-0043, Japan} 

\abst{We study the spin-fluctuation-mediated $s\pm$-wave superconductivity in the bilayer Hubbard model with vertical and diagonal interlayer hoppings. As in the two-leg ladder model with diagonal hoppings, studied previously by the present authors, superconductivity is strongly enhanced when one of the bands lies just below (or touches) the Fermi level, that is, when the band is incipient. The strong enhancement of superconductivity is because large weight of the spin fluctuations lies in an appropriate energy range, whereas the low energy, pair-breaking spin fluctuations are suppressed. The optimized eigenvalue of the linearized Eliashberg equation, a measure for the strength of superconductivity, is not strongly affected by the bare width of the incipient band, but the parameter regime where superconductivity is optimized is wide when the incipient band is narrow, and in this sense, the coexistence of narrow and wide bands is favorable for superconductivity.}

\begin{document}
\maketitle

\section{Introduction}

In many of the iron-based superconductors, electron-like and hole-like Fermi surfaces coexist, and this has led to the scenario of spin-fluctuation mediated $s\pm$ pairing, where the nesting between the two Fermi surfaces give rise to spin fluctuations, which acts as a pairing glue\cite{Hirschfeldrev,KKrev1,Chubukovrev,KKrev2,MazinSingh,KKPRL}. On the other hand, in some iron-based superconductors, the hole Fermi surface is found to sink below the Fermi surface, but still gives rather high $T_c$\cite{Iimura,KFe2Se2,KFe2Se2ARPES,STO,STO2,XJZhou,Takahashi,Borisenko,Ding,LiOH}. This has led to theoretical studies on the role of ``incipient band'', a band sitting just below (or above) the Fermi level, played in the occurrence of superconductivity\cite{DHLee,Hirschfeld,Hirschfeldrev,YBang,YBang2,YBang3,Borisenko,Ding}. In fact, it was shown earlier in ref.\cite{Kuroki}, for a model for the two-leg ladder-type cuprates, that the interband scattering processes between a wide band that intersects the Fermi level and a narrow band just below the Fermi level can give rise to an strong enhancement of spin-fluctuation-mediated superconductivity\cite{Kuroki}. More recently, this theory has been extended to various quasi-one-dimensional lattice structures\cite{Matsumoto}, and also, it has been pointed out that a similar two-leg-ladder-like electronic structure is hidden in the Ruddlesden Popper bilayer compound\cite{Ogura,OguraDthesis}. Other models with coexisting wide and flat bands have also been pointed out to enhance superconductivity\cite{KobayashiAoki,Misumi,Sayyad,Aokireview}.  Also in the bilayer Hubbard model\cite{Bulut,KA,Scalettar,Hanke,Santos,Mazin,Kancharla,Bouadim,Fabrizio,Zhai,Maier,MaierScalapino,Nakata,MaierScalapino2}, which can be considered as a two-dimensional version of the two-leg ladder Hubbard model, the effect of the incipient band has also been studied \cite{MaierScalapino,Nakata,MaierScalapino2}. There it was also found that $s\pm$-wave superconductivity is strongly enhanced when the edge of one of the bands sit close to the Fermi level, namely, when one of the bands is nearly incipient. In refs.\cite{MaierScalapino,Nakata}, it was revealed that superconductivity is enhanced when the spin fluctuations have large weight at finite energies, namely, when the band is incipient, (most of) the band is below (or above) the Fermi level, so that the interband interaction leads to development of spin fluctuations at finite energies, while the low energy (near-zero-energy) spin fluctuations are suppressed.

In the present study, we study the bilayer Hubbard model, where we consider not only the vertical interlayer hopping, but also the diagonal interlayer ones (Fig.\ref{fig1} upper left), which makes the bonding band narrow and the antibonding band wide, as in the two-leg ladder with diagonal hoppings (Fig.\ref{fig1} upper right)\cite{Kuroki,Matsumoto}. We compare the results to those obtained for the two-leg ladder, and discuss what is the key factor in the enhancement of superconductivity owing to the presence of the incipient band.

\section{The models and methods}

\begin{figure}
\includegraphics[width=7.5cm]{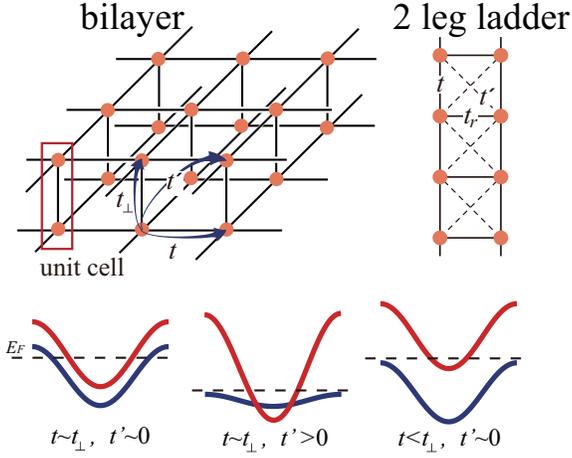}
\caption{Upper panel : the bilayer lattice (left) and the two-leg ladder lattice (right). Lower panels : schematic images of the bonding and antibonding bands of the bilayer lattice. Left : $t\sim t_\perp$, $t'\sim 0$, middle : $t\sim t_\perp$, $t'> 0$, right : $t< t_\perp$, $t'\sim 0$. In the middle and right panels, the bonding band is incipient. } 
\label{fig1}
\end{figure}

The bilayer lattice shown in Fig.\ref{fig1} is given, in standard notation, as 
\begin{eqnarray}
H&=&t\sum_{\alpha=1,2}\sum_{\sigma=\uparrow,\downarrow}\sum_{\langle i,j\rangle}
(c_{i\sigma\alpha}^\dagger c_{j\sigma\alpha}+H.c.)\nonumber\\
&+&t_\perp\sum_{\sigma=\uparrow,\downarrow}
(c_{i\sigma1}^\dagger c_{i\sigma2}+H.c.)\nonumber\\
&+&t'\sum_{\sigma=\uparrow,\downarrow}\sum_{\langle i,j\rangle}
(c_{i\sigma1}^\dagger c_{j\sigma2}+H.c.),
\end{eqnarray}
where $i,j$ specify unit cells (pairs of sites connected by the vertical hopping $t_\perp>0$), $\langle i,j\rangle$ denotes nearest neighbor unit cells, and $\alpha=1,2$ specifies the layers. $t'$ is the diagonal interlayer hopping, and the intralayer nearest neighbor hopping $t=1$ is taken as the unit of energy. 

In momentum space, the bonding and antibonding bands are given as 
\begin{eqnarray}
\varepsilon_b(k_x,k_y)&=&2(t-t')(\cos(k_x)+\cos(k_y))-t_{\perp}\\
\varepsilon_{ab}(k_x,k_y)&=&2(t+t')(\cos(k_x)+\cos(k_y))+t_{\perp}.
\end{eqnarray}
We consider $t'$ in the range $0\leq t'\leq 1$ ; at $t'=0$ the bonding and antibonding bands have the same width, and for $0<t'\leq 1$ the bonding band is narrower than the antibonding one (see the bottom panels of Fig.\ref{fig1}). Especially at $t'=1$, the bonding band is perfectly flat. The band filling $n$ is defined as the average number of electrons per unit cell; $n=2$ corresponds to half filling. We  focus on cases with $n>2$ because we are interested in the situation where the narrow bonding band is made (nearly) incipient by raising the Fermi level up to its top (see the bottom panels of Fig.\ref{fig1})\cite{comment}. Note that the parameter regime considered in the present study is equivalent to that with $0\geq t'\geq -1 $ and $n<2$, as can be seen by electron-hole transformation. 

On top of this tightbinding model, we consider the on-site repulsive Hubbard interaction term,
\begin{equation}
H_{\rm int}=U\sum_i\sum_{\alpha=1,2} n_{i\alpha\uparrow}n_{i\alpha\downarrow},
\end{equation}
where $n_{i\alpha\sigma}$ is the number operator of electrons with spin $\sigma$ at the $i$-th unit cell, layer $\alpha$.  Unless noted otherwise, $U=6$ is adopted, which is a typical value (in units of $t$) for the cuprates and related transition metal oxides\cite{Vaugier_2012_cRPA_3d-4d,Mravlje_2011_cRPA,Jang_2016_cRPA}. We apply the fluctuation  exchange (FLEX) approximation\cite{Bickers,Dahm} to obtain the renormalized Green's function. Namely, bubble and ladder type diagrams are collected to obtain the spin and charge susceptibilities, which enter the effective interaction that is necessary to obtain the self energy. The Dyson's equation is solved using the self energy, which gives the renewed Green's function, and the self energy is recalculated. This iteration process is repeated till convergence is attained. Green's function is first obtained in the site representation, namely, in the form of $G_{\alpha\beta}$, where $\alpha$, $\beta$ denotes the sites within a unit cell. Then it is transformed into the band representation by a unitary transformation. The absolute value of Green's function at the lowest Matsubara frequency $|G(\Vec{k},i\pi k_BT)|$ is used to represent the Fermi surface of the renormalized bands. We also calculate the imaginary part of the dynamical spin susceptibility $\chi(\bm{q},\omega)$. $\chi(\bm{q},\omega)$ is obtained by Pad\'{e} analytical continuation of the FLEX spin susceptibility obtained within the Matsubara formalism. 
As a quantity that measures the strength of the spin fluctuation, we define Im $\Gamma(\omega)$ as  
\begin{equation}
\sum_{\bm{q}}\mathrm{Im\,}\chi(\bm{q},\omega)\equiv \mathrm{Im\,}\Gamma(\omega).
\end{equation}

To study superconductivity mediated by the spin fluctuation, the linearized Eliashberg equation, 
	\begin{eqnarray}
	\lambda \Delta_{ll^{\prime}}(k) &=& -\frac{T}{N}\sum_{k^{\prime}m_i}\Gamma_{lm_1 m_4l^{\prime}}(k-k^{\prime})G_{m_1m_2}(k^{\prime})\nonumber\\
	&&\times \Delta_{m_2m_3}(k^{\prime}) G_{m_4m_3}(-k^{\prime})
	\end{eqnarray}	
is solved, where $k$ stands for a combination of the wave vector and the Matsubara frequency, the subscripts denote the sites within a unit cell, $T$ is the temperature, $N$ is the number of $k$-point mesh,  $\Delta$ is the anomalous self energy, $\Gamma$ is the pairing interaction, whose main contribution comes from the FLEX spin susceptibility mentioned above. 
The eigenvalue $\lambda$ of the linearized Eliashberg equation reaches unity at the superconducting transition temperature $T = T_c$, so that when it is calculated at a fixed temperature, systems with higher $T_c$ give larger eigenvalues. In other words, $\lambda$ calculated at a fixed temperature can be considered as a measure of $T_c$. Throughout the study, we calculate $\lambda$ at $T=0.05$. Within the parameter regime studied, $\Delta$ that gives the largest $\lambda$ is always found to be of the $s\pm$-wave type, where $\Delta$, when transformed into band representation, has nodeless $s$-wave symmetry, and changes its sign between bonding and antibonding bands\cite{KA}. We will call this $s\pm$-wave pairing even when the bonding band does not intersect the Fermi level.  As for the band filling, we restrict ourselves to $n\geq 2.1$, since it is difficult to treat band fillings too close  to half-filling $(n=2)$ within FLEX\cite{comment2}.   In the calculation, we take $32\times 32$ two-dimensional $k$-point mesh and 1024 Matsubara frequencies. 

In some cases, we compare the results for the bilayer model with those for the Hubbard model on a two-leg ladder lattice (Fig.\ref{fig1} upper right), which was partially studied in ref.\cite{Matsumoto} in a similar way. The two-leg ladder can be considered as a one-dimensional counterpart of the bilayer lattice, where $t_\perp$ is replaced by $t_r$, the nearest neighbor hopping in the rung direction. (In ref.\cite{Matsumoto}, only the case of $t_r=t$ was studied.) Also, we introduce small interladder hoppings $t_i=0.1$ so as to make the system quasi-one-dimensional, as in ref.\cite{Matsumoto}.

\section{Cases when the interlayer vertical hopping is equal to the intralayer ones}
\label{tperpeq1}
\begin{figure}
\includegraphics[width=7cm]{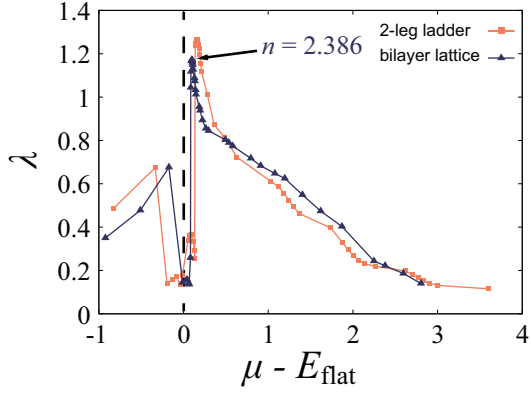}
\centering
\caption{Eigenvalue of the Eliashberg equation $\lambda$ at $T=0.05$ plotted against the bare Fermi level measured from the flat band energy for the case of $t'=1$, $t_\perp=1$. The case for the two-leg ladder\cite{Matsumoto} is shown for comparison.}
\label{fig2}
\end{figure}

In this section, we concentrate on the cases when 
the interlayer vertical hopping $t_\perp $ is equal to the intralayer nearest neighbor 
hoppings $t$, taken as the unit of the energy.
We start with the case when the bonding band is perfectly flat, namely, when $t'=1$.
In Fig.\ref{fig2}, we plot the eigenvalue of the Eliashberg equation $\lambda$  
as functions of the bare Fermi level measured from the flat band energy for 
both the bilayer and the two-leg ladder lattices. 
As was already seen for the two-leg ladder in ref.\cite{Matsumoto}, 
the eigenvalue for the bilayer lattice also exhibits a sharp maximum 
when the Fermi level comes close to the flat band energy, but decreases 
rapidly when it is too close to the flat band. Interestingly, the 
two models exhibit very similar dependencies against the Fermi level.  
Here we stress that the maximum value of $\lambda$ is very large ; it largely exceeds unity at the temperature of $T=0.05$, and the $T_c$ is actually close to $T=0.1$. This implies strong enhancement of superconductivity compared to the case of the single layer Hubbard model on a square lattice, a model for the high $T_c$ cuprates, where the typical $T_c$ is $0.02-0.03t$\cite{Bickers}.

\begin{figure}
\includegraphics[width=7cm]{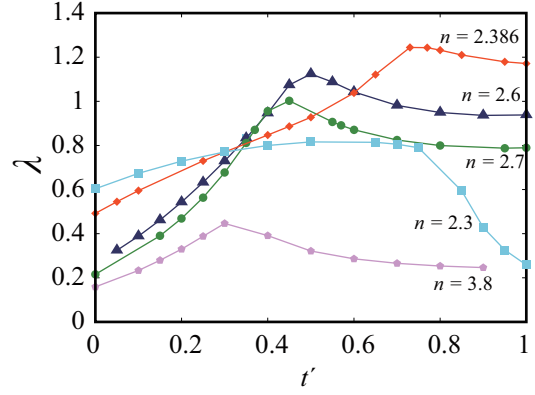}
\centering
\caption{$\lambda$ at $T=0.05$ plotted against $t'$ for various band fillings.}
\label{fig3}
\end{figure}

Next we turn to the cases when the bonding band has finite band width.
In Fig.\ref{fig3}, we plot, for various band fillings $n$, 
the eigenvalue against $t'$, which controls the band width of each band.
For each $n$, the eigenvalue $\lambda$ is maximized at a certain $t'$, and the $t'$ value which maximizes $\lambda$ is smaller for larger $n$. This 
variation of $\lambda$ against $t'$ resembles  that seen for the two-leg ladder \cite{Matsumoto} (see Fig.9 of ref.\cite{Matsumoto});
there it was revealed that the finite band width of the bonding band 
brings its edge closer to the Fermi level, which leads to the enhancement of 
superconductivity, but when the bonding band edge comes too close to, or 
intersects the Fermi level, superconductivity is degraded due to strong renormalization effects. Namely, superconductivity 
is optimized when the bonding band is incipient. The value of $t'$ at 
which the bonding band touches the Fermi level is smaller (requires larger 
band width) for larger $n$ because the Fermi level is more raised.

\begin{figure}
\includegraphics[width=7cm]{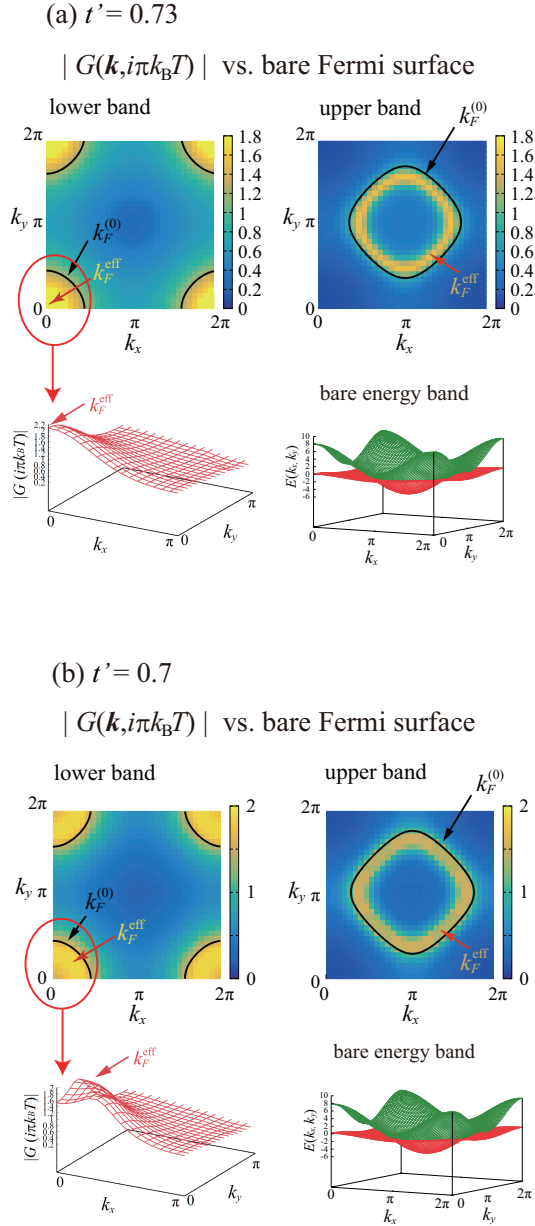}
\centering
\caption{$|G(\Vec{k},i\pi k_BT)|$ for the two bands, the bare Fermi surface, and the bare energy band plotted for (a) $t'=0.73$ and $t'=0.7$ for the band filling $n=2.386$. $k_F^{(0)}$ is the bare Fermi surface, while $k_F^{eff}$ denote the (local) maximum of $|G(\Vec{k},i\pi k_BT)|$, which corresponds to the Fermi surface of the renormalized band when the band intersects the Fermi level. Note that the upper (lower) band implies the green (red) portion shown in the bare band ; it is different from the bonding and antibonding bands depicted as blue and red bands in Fig.\ref{fig1}. However, the Fermi surface and hence the ridge of $|G(\Vec{k},i\pi k_BT)|$ of the antibonding (bonding) band are the same as those of the upper (lower) band.}
\label{fig4}
\end{figure}

To show that a similar situation takes place in 2D, we concentrate on the 
case of $n=2.386$, for which $\lambda$ is maximized at $t'=1$, and analyze Green's function. In Fig.\ref{fig4}, we show Green's function for two cases: 
$t'=0.73$, where $\lambda$ is maximized, and $t'=0.7$, where $\lambda$ 
is somewhat degraded compared to the optimal value. For $t'=0.73$, 
Green's function takes its maximum at the wave vector (0,0), where the 
lower (bonding) band takes its maximum energy value. This means that the bonding band is below (or just touches) the Fermi level. For $t'=0.7$, on the other hand, Green's function of the lower (bonding) band exhibits a ``ridge'' structure, which corresponds to a 
Fermi surface with finite size. This indicates that $\lambda$ takes its maximum 
when the bonding band is incipient,
and when the bonding band intersects the Fermi level and a Fermi surface 
is formed, superconductivity starts to be degraded.

So the tendency between the two-leg ladder and the bilayer lattice 
is again similar. Actually, however, 
this may be surprising considering the large difference 
between one dimension(1D) and 2D in the density of states (DOS) at the band edge. In ref.\cite{Matsumoto}, the present authors interpreted that in 1D, the DOS is diverging at the band edge, so that it plays a role similar to that of the flat band, and hence high $T_c$ is obtained even when the bonding band has finite 
band width. The present result for the 2D bilayer lattice shows that 
a divergingly large DOS {\it at the band edge} is not necessary for strongly enhanced superconductivity. 

We note that, strictly speaking, there are differences between the bilayer and the two-leg ladder in the variance of $\lambda$ against $t'$. In the present bilayer case, for all the band fillings studied, the bonding band with some finite band width gives higher $\lambda$ than when it is perfectly flat, while in the two-leg ladder case, for the band filling that gives the largest $\lambda$ for the flat band case, introduction of finite band width leads to a reduction of $\lambda$ (Fig.9 of ref.\cite{Matsumoto}).  Another difference is that in the present bilayer case, there is a cusp in the $\lambda$ variation when the bonding band touches the Fermi level, followed by a rapid decrease of $\lambda$ as the bonding band firmly forms a Fermi surface, but in the two-leg ladder case (Fig.9 of ref.\cite{Matsumoto}), $\lambda$ smoothly varies against $t'$, and the suppression of $\lambda$ after the bonding band forms a Fermi surface is mild.  We will also come back to the origin of this difference in the Discussion section.

\section{Cases when the vertical interlayer hopping is larger than the intralayer ones}
\label{tperpgt1}
\begin{figure}
\centering
\includegraphics[width=7cm] {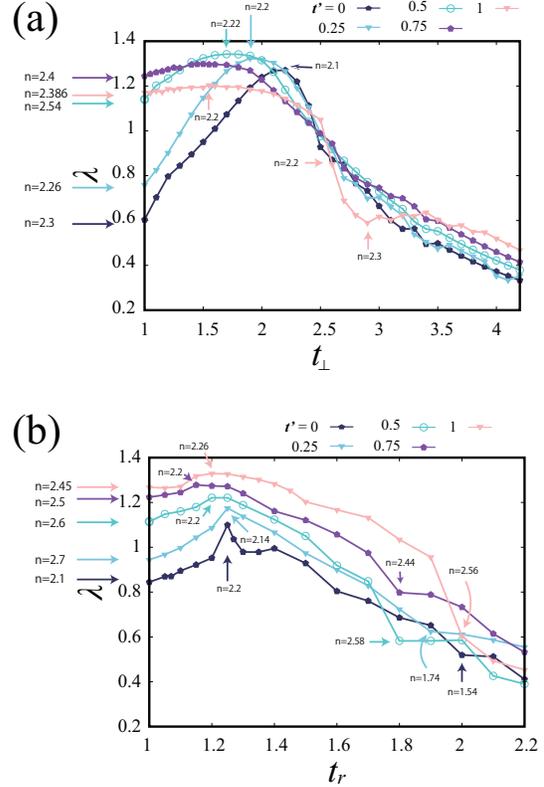}
\caption{(a) Maximized $\lambda$ of the bilayer Hubbard model at $T=0.05$ plotted against $t_{\perp}$ for various $t'$. $\lambda$ is maximized for each set of $(t',t_{\perp})$ by varying the band filling $n$. At some points, $n$ that maximizes $\lambda$ is denoted by arrows. (b) Similar plot for the two-leg Hubbard ladder model, where $t_\perp$ is replaced by $t_r$.}
\label{fig5}
\end{figure}

In this section, we consider the cases when the vertical interlayer 
hopping $t_\perp$ is larger 
than the intralayer nearest neighbor hoppings $t$. 
This is motivated by 
previous studies which show that high $T_c$ is realized in the bilayer lattice model without the diagonal hopping when the vertical hopping becomes appropriately large\cite{Bulut,KA,Maier,Nakata,MaierScalapino}. From the band picture viewpoint, this enhancement of superconductivity can be understood as a consequence of the bonding band made incipient by increasing $t_\perp$ up to an appropriate value (see the lower right panel of Fig.\ref{fig1}). Here, for each $t'$, we vary $t_\perp$ as the horizontal axis, and for each combination of $(t',t_\perp)$ we vary the band filling within $n\geq 2.1$ to maximize the eigenvalue $\lambda$. The result is shown in Fig.\ref{fig5}. We have confirmed (as in the previous section) that the bonding band is (nearly) incipient in cases where $\lambda$ exceeds unity, but when $\lambda$ is small as in the case of $t'\sim 0$ and $t_\perp\sim 1$, too large $n$ is required to make the bonding band incipient, so instead $\lambda$ is maximized at a band filling close to half filing, where the bonding band is not incipient. From the present result, one can see that the maximum value of $\lambda$ itself does not depend so much on $t'$, that is, the bonding band width.

\begin{figure}
\centering
\includegraphics[width=7cm] {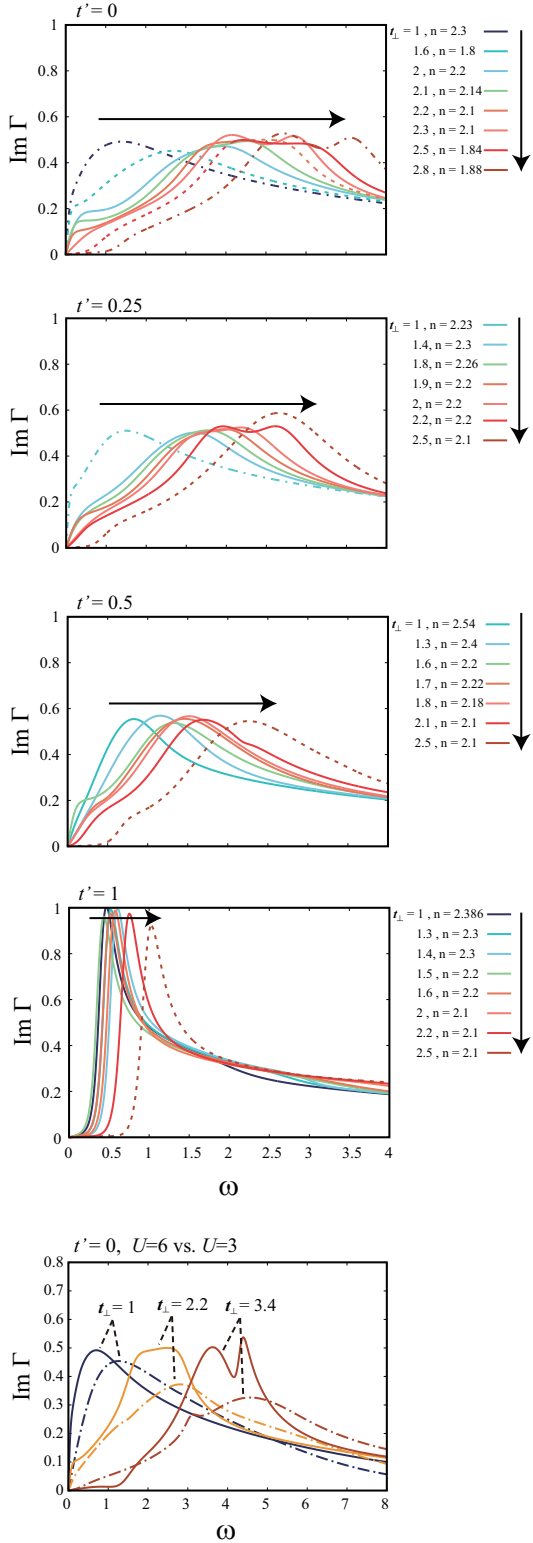}
\caption{Upper four panels : Im$\Gamma$ plotted against $\omega$ for various $t'$ and $(t_\perp,n)$. The parameter sets are chosen from those adopted in Fig.\ref{fig5}. The solid, dashed, and dash-dotted lines correspond to the cases where $\lambda>1.1$, $1.1>\lambda>0.8$, and $0.8>\lambda$, respectively, in Fig.\ref{fig5}. Bottom panel : Im$\Gamma$ plotted against $\omega$ for $t'=0$ and $U=6$ (solid) or $U=3$ (dash-dotted). $(t_\perp,n)$ are chosen from those adopted in Fig.\ref{fig8}, where three values of $t_\perp$ are chosen ; $t_\perp=1$, $t_\perp=2.2$ (where the bonding band for $U=6$ sinks just below the Fermi level) and $t_\perp=3.4$ (where the bonding band for $U=3$ sinks just below the Fermi level).}
\label{fig6}
\end{figure}

On the other hand, $\lambda$ takes large values only in a limited $t_\perp$ regime when $t'$ is small, while it remains to take large values in a wide range of $t_\perp$ (if the band filling is optimized) when $t'$ is large.  This can be explained as follows. When $t'$ is small, the bonding band and the antibonding band have similar band width, so that when $t_\perp$ is small (i.e., when the level offset between the two bands is small, so that the energy level of the bonding band with respect to that of the antibonding band is high), large amount of electrons is necessary in order to make the bonding band sink below the Fermi level. In such a case, the band filling is too far away from half filling, which is unfavorable for superconductivity. By contrast, when $t'$ is large, the bonding band is narrow so that it can be made incipient for band fillings not so far away from half filling even when $t_\perp$ is not so large. Furthermore, even when there is a bare Fermi surface, if the band edge if close to the Fermi level, there is a tendency that the electron correlation effects make the band even more close to the incipient situation (see Fig.\ref{fig4}(a), ref.\cite{Matsumoto}, and also Fig.\ref{fig9}(a)(b) in the Discussion section), so that when the bonding band is narrow, the range of the parameter regime where it becomes nearly incipient is even further widened by the correlation effects.

For comparison, we also show in Fig.\ref{fig5}(b) a similar plot for the two-leg ladder, where $t_\perp$ is replaced by $t_r$, the nearest neighbor hopping in the rung direction  (see Fig.\ref{fig1}). The result is basically similar to that for the bilayer model, and the maximum values of $\lambda$ are also similar.  However, if we look more closely, we see that the $t'=1$ case with a perfectly flat bonding band exhibits the largest $\lambda$ in most of the $t_r$ regime. This is a consequence of what we mentioned in the end of the previous section, i.e., for the two-leg ladder, the flat bonding band is favorable compared to the case with finite bonding band width. Also, we find that plots similar to Fig.\ref{fig5}(a)(b) for a smaller $U=3$ are significantly different between the bilayer and two-leg ladder models. We will come back to these points in the Discussion section.

\section{Role of spin fluctuations in various energy ranges}

The important role played by finite energy spin fluctuations in the enhancement of superconductivity has been pointed out for the bilayer Hubbard model without diagonal hoppings in previous studies\cite{MaierScalapino,Nakata}. In Fig.\ref{fig6}, we plot the the $q$-space summation of the imaginary part of the dynamical spin susceptibility, Im$\Gamma$,  as functions of the frequency $\omega$ for various $t'$ and $(t_\perp,n)$. The parameter sets are chosen from those adopted in Fig.\ref{fig5}. The solid lines correspond to cases where the eigenvalue of the Eliashberg equation is large. It can be seen that when the low energy part ($\omega < O(0.1)$) of the spin fluctuation is large (this is when the Fermi surface of the bonding band is firmly formed), superconductivity is degraded. This is because the strong low energy spin fluctuations due to relatively good Fermi surface nesting strongly renormalizes the quasiparticles and hence are pair breaking\cite{Millis,Matsumoto}.  Large values of $\lambda$ is obtained when the low energy part ($\omega < \sim 0.1$) of the spin fluctuation is suppressed while the spin fluctuation in the range $\sim 0.1<\omega<\sim 1$ remains appreciable. The latter is considered to be the frequency range most effective as a pairing glue. Hereafter, we will call the spin fluctuations in this energy range ``pairing effective''. When the spin fluctuation weight is transferred to too high energies, superconductivity is once again degraded.

\begin{figure}
\includegraphics[width=8cm]{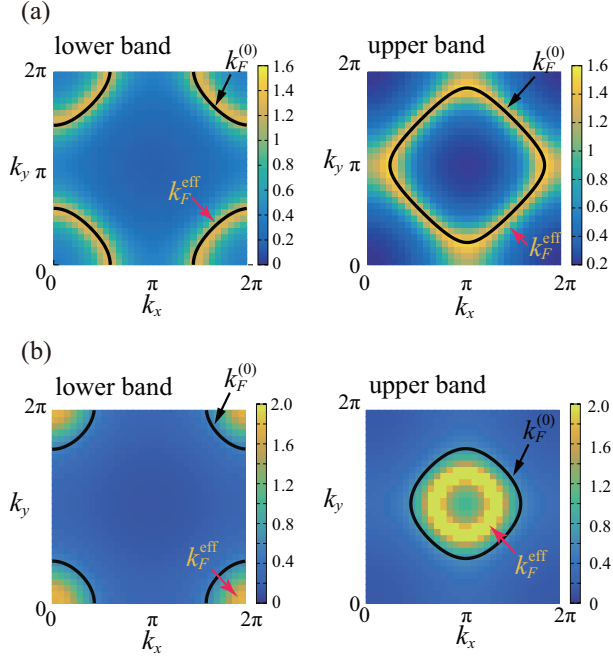}
\centering
\caption{$|G(\Vec{k},i\pi k_BT)|$ for the two bands and the bare Fermi surface plotted for (a) $t'=0$, $t_\perp=1$, $n=2.3$ and (b)  $t'=0$, $t_\perp=2.2$, $n=2.1$. $k_F^{(0)}$ and $k_F^{eff}$ are similar to those in Fig.\ref{fig4}.}
\label{fig7}
\end{figure}

To show the difference in the quasiparticle renormalization between cases when the bonding band is incipient and when it firmly forms a Fermi surface, in Fig.\ref{fig7}, we compare Green's function between the two cases with $t'=0$. For $t_\perp$=2.2 and $n=2.1$, where the bonding band is nearly incipient and $\lambda$ is large, $|G(\Vec{k},i\pi k_BT)|$ of the upper (antibonding) band exhibits a sharp ridge, meaning that the quasiparticle renormalization is weak. By contrast, for $t_\perp=1$ and $n=2.3$, where the Fermi surface is firmly formed and $\lambda$ is small,  $|G(\Vec{k},i\pi k_BT)|$ of the upper (antibonding) band is more suppressed due to stronger renormalization. Another difference between the two cases is that for $t_\perp$=2.2 and $n=2.1$, the bonding band is made nearly incipient through the electron-electron interaction effect (the bare Fermi surface is not incipient), while the electron-electron interaction barely affects the Fermi surface in the case of $t_\perp=1$ and $n=2.3$.

\section{Discussion}
So far we have seen that the incipient band situation is favorable for superconductivity in that (i) the low energy spin fluctuations (nearly zero, less than $0.1t$), which have pair-breaking effect through quasiparticle renormalization, are suppressed, and (ii) the moderate energy spin fluctuations ($\sim 0.1t-t$), effective as pairing glue for superconductivity, develop. From this viewpoint, we further discuss some issues in this section. 

\begin{figure}
\includegraphics[width=8cm]{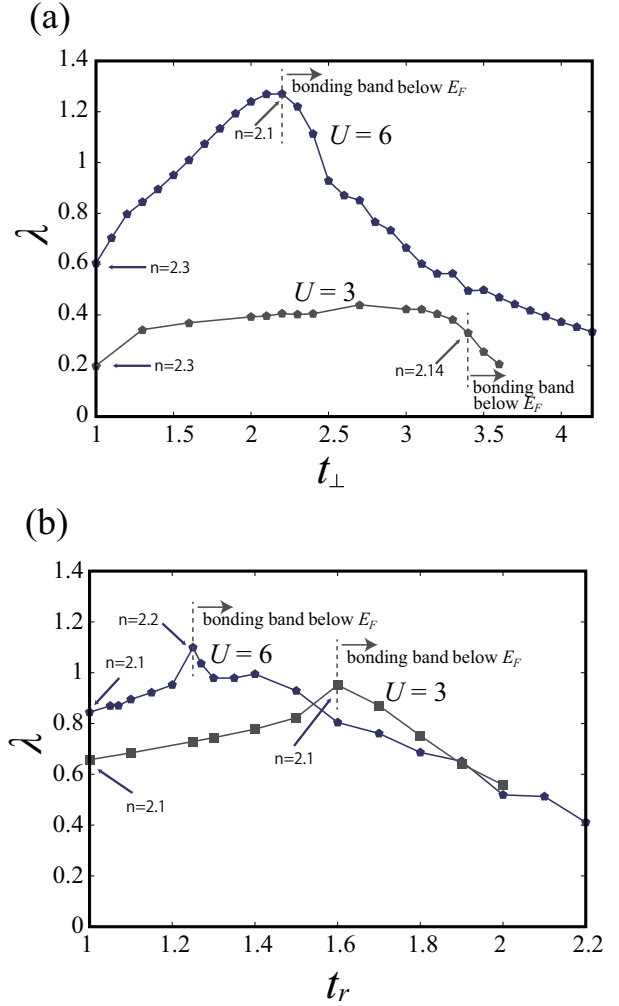}
\centering
\caption{(a) The result for $t'=0$ is extracted from Fig.\ref{fig5}(a), and compared to that of $U=3$, $t'=0$. (b) A similar plot for the two-leg ladder, where the result for $t'=0$ is extracted from Fig.\ref{fig5}(b), and compared to that of $U=3$, $t'=0$.}
\label{fig8}
\end{figure}

\begin{figure}
\includegraphics[width=8cm]{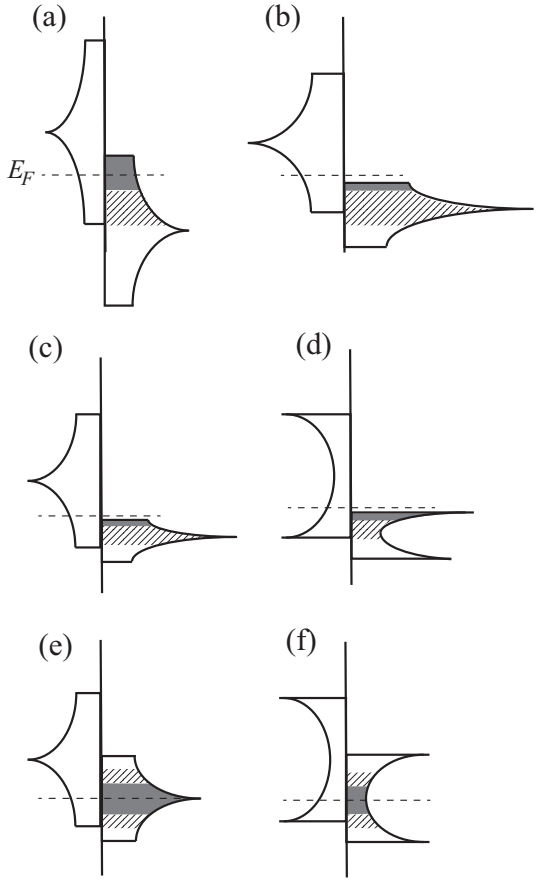}
\centering
\caption{Schematic images of the renormalized DOS of the bilayer and two-leg ladder models. In each figure, the left (right) side of the vertical line depicts the DOS of the antibonding (bonding) band. The gray area denotes the portion of the bonding band DOS which gives rise to the low energy (pair-breaking) spin fluctuations, and the hatched area is the portion of the bonding band DOS contributing to the spin fluctuations that are effective for superconductivity. (a)(b)(c)(e) are for the bilayer model, and (d)(f) are for the two-leg ladder model. (a) small $U$ case, and (b) large $U$ case with the bonding band being incipient. (c) and (d) are cases where the bonding band is incipient. (e) and (f) are cases where $t'\sim 0$ and $t_\perp\sim 1$, so the bonding band intersects the Fermi level (large amount of electrons are  required for the bonding band to be incipient). }
\label{fig9}
\end{figure}

The first issue is the effect of electron correlation on the band width. In Fig.\ref{fig8}(a), we compare between $U=6$ and $U=3$ a plot similar to Fig.\ref{fig5}(a) for $t'=0$. $\lambda$ is strongly suppressed for $U=3$ compared to $U=6$. Also, much larger $t_\perp$ is required for the bonding band to sink below the Fermi level, where $\lambda$ is further suppressed. The result indicates that the incipient band situation in the bilayer model is not so favorable for superconductivity when $U$ is small. In the bottom panel of Fig.\ref{fig6}, we also compare Im$\Gamma$ for the two values of $U$. For $U=3$, not only Im$\Gamma$ is reduced, but also it is distributed in a wide $\omega$ range. Conversely, the spin fluctuation weight is ``squeezed'' into a narrow energy range regime due to the electron correlation effect when $U$ is large.  This can be understood as follows. In ref.\cite{OguraDthesis}, it was revealed using FLEX that the portion of the bonding band close to the Fermi level is strongly renormalized due to electron correlation to give a DOS schematically depicted in Fig.\ref{fig9}(b). This effect is also seen in a dynamic cluster quantum Monte Carlo study for the bilayer Hubbard model without $t'$\cite{MaierScalapino2}. Due to this effect, smaller $t'$ and/or $t_\perp$ would suffice for the bonding band to become incipient, and also the spin fluctuation spectrum is squeezed into a narrower frequency regime, so that more of its weight lies within the pairing-effective energy range  when the bonding band is incipient. The present view is further confirmed from a similar plot of $\lambda$ for the two-leg ladder (Fig.\ref{fig8}(b)), where even for $U=3$, when the bonding band is incipient, $\lambda$ takes large values similar to those for $U=6$. Here, the density of states is concentrated in the energy range close to the Fermi level even in the absence of the electron correlation due to quasi-one-dimensionality, so that the incipient band is favorable for superconductivity even for small $U$.   

The second issue is on the difference in the bonding band width dependence between the bilayer and two-leg ladder models, which was mentioned in the end of sections \ref{tperpeq1} and \ref{tperpgt1}. We interpret this in terms of the relation between the frequency dependence of the spin fluctuations and the shape of the DOS. Since we consider the situation where the antibonding band intersects the Fermi level, the portion of the bonding band that is about $\omega$ away from the Fermi level mainly contributes to the spin fluctuations having the frequency $\omega$. Let us start by considering the case when the bonding band is perfectly flat. In this case, regardless of bilayer or two-leg ladder, if the flat band lies below but close to the Fermi level, all the states in the bonding band will contribute to the low energy, pair-breaking spin fluctuations as well as to the pairing-effective spin fluctuations. In other words, the number of states that can contribute to the pairing-effective and pair-breaking spin fluctuations are the same. Now we compare this with the cases when the bonding band has finite band width. When the bonding band is just below the Fermi level, around the top of the bonding band contributes to the pair-breaking spin fluctuations, while the portion of the band somewhat away from the top contributes more to the pairing-effective ones.  Therefore, for the bilayer model (Fig.\ref{fig9}(c)), the DOS of the portion that produces the pairing-effective spin fluctuations is larger than the DOS of the portion that gives rise to the pair-breaking ones. This is the reason why finite bonding band widths gives larger optimized $\lambda$ than that for a perfectly flat one in the bilayer model. By contrast, in the two-leg ladder case, where the DOS at the bonding band top is (nearly) diverging (Fig.\ref{fig9}(d)), the DOS of the portion contributing to the pair-breaking spin fluctuations is large, while the pairing-effective spin fluctuations originate from the portion of the band with smaller DOS. Hence, for the two-leg ladder, the flat bonding band case, where the number of states that can contribute to the pairing-effective and pair-breaking spin fluctuations are the same, is the best for superconductivity provided that the band filling is optimized. 

A similar consideration also explains the reason why superconductivity in the bilayer model is rapidly suppressed after $t'$ is reduced enough for the bonding band to intersect the Fermi level (Fig.\ref{fig3}). Namely, the DOS at the Fermi level increases as the van Hove singularity of the bonding band approaches the Fermi level (Fig.\ref{fig9}(e)), so that the pair breaking low energy spin fluctuations quickly develop (see the dash-dotted lines in Fig.\ref{fig6}). This is in contrast to the two-leg ladder case, where the DOS at the Fermi level {\it decreases} after the Fermi surface of the bonding band is formed (Fig.\ref{fig9}(f)), leading to a milder suppression of superconductivity (Fig.9 of ref.\cite{Matsumoto}). 

The FLEX approximation adopted in the present study is a weak-coupling approach, whose reliability in the large $U$ regime is not so clear. As an alternative and complementary approach, quite recently, one of the present authors and his coworker performed a multivariable variational Monte Carlo study on the two-leg ladder and the bilayer Hubbard models, and have obtained numerical results which support the present conclusion. This study will be published elsewhere\cite{Kato}.

Finally, let us discuss the relevance of the present view to the existing superconductors. As mentioned in the Introduction, the importance of (nearly) incipient bands in the iron-based superconductors has already been pointed out. In fact, our FLEX calculation on a realistic five orbital model for a 1111-type iron-based superconductor have shown that superconductivity is optimized when the $d_{xy}$ hole Fermi band is incipient\cite{Nakata}. Also mentioned in the Introduction, our original motivation came from the electronic structure of the ladder cuprates\cite{Kuroki}. Quite recently, one of the present authors and his coworker revisited this problem using a realistic model Hamiltonian of the ladder-type cuprates derived from first principles calculation\cite{Sakamoto}.  Actually, the present view might even have some relevance to the ordinary cuprates with CuO$_2$ planes, if we interpret the extended van Hove singularity observed in photoemission studies somewhat below the Fermi level\cite{Gofron} as corresponding to the ``incipient band'', although this picture obviously cannot be straightforwardly accepted because the ordinary cuprates are single band systems.

\section{Conclusion}

We have studied the spin-fluctuation-mediated $s\pm$-wave superconductivity in the bilayer Hubbard model with vertical and diagonal interlayer hoppings within the fluctuation exchange approximation. Superconductivity is strongly enhanced when  one of the bands (the bonding band here) is nearly incipient. This tendency is quite similar to that found in the two-leg ladder Hubbard model with diagonal hoppings\cite{Matsumoto}. The origin of the strong enhancement of superconductivity is that when the bonding band is nearly incipient, large weight of the spin fluctuation lies in a pairing-effective regime appropriate for high $T_c$ superconductivity. When the bonding band firmly forms a Fermi surface, the low energy spin fluctuations strongly develop, which leads to strong renormalization of the quasiparticles and hence suppression of superconductivity\cite{Millis}, while when the bonding band is too far away from the Fermi level, the spin fluctuation weight is transferred to too high energies, which cannot be exploited as an effective pairing glue. 

The dimensionality of the lattice (bilayer or two-leg ladder) or the bare width of the incipient band does not strongly affect the maximum value of the eigenvalue of the Eliashberg equation when the electron-electron interaction is large enough, which implies that a flatness of the incipient band is not a prerequisite for the strong enhancement of superconductivity. On the other hand, the renormalization of the band due to correlation effects, which enhances the DOS of the incipient band\cite{OguraDthesis,MaierScalapino2}, is favorable for superconductivity because such an effect would enhance the weight of pairing-effective spin fluctuations. Also, when the bonding band is narrow, the incipient band situation and hence the strong enhancement of superconductivity is realized in a wide parameter regime. In this sense, the coexistence of wide and narrow bands is favorable for superconductivity. 
\ \\

\begin{acknowledgment}
\acknowledgment
We thank Daichi Kato, Masayuki Ochi, Shungo Nakanishi, Hidetomo Usui, and Hideo Aoki for valuable discussions. This study is supported by JSPS KAKENHI Grant Number JP18H01860.

\end{acknowledgment}


\begin{thebibliography}{99}
\bibitem{comment} We can also make one of the bands incipient in the $n<2$ regime by lowering the Fermi level down to the band bottom, but in that case, when $t'>0$ is large, the band filling is too far away from the Fermi level, so that the electron correlation, and hence the enhancement of superconductivity, are weak.
\bibitem{comment2} We have also investigated the regime $n\leq 1.9$, but $\lambda$ is found to be maximized for $n \geq 2.1$. For $t'=0$, the band filling $n$ and $4-n$ give identical results within numerical error due to electron-hole symmetry.
\end{thebibliography}
\end{document}